%% file: 2026_Time-Glide_arxiv.tex
\documentclass{nature}

\usepackage[utf8]{inputenc}
\usepackage[T1]{fontenc}
\usepackage[letterpaper,margin=1in]{geometry}
\usepackage{graphicx}
\usepackage{amsmath,amssymb}
\usepackage{bm}
\usepackage{xcolor}
\usepackage{booktabs}
\usepackage{caption}
\usepackage{placeins}
\usepackage{hyperref}
\hypersetup{
  pdftitle={Temporal glide symmetry enforces a parity sideband selection rule in scalar bulk media},
  hidelinks
}

\graphicspath{{figures/}}
\spacing{1.5}

\let\naturegraphicxincludegraphics\includegraphics
\AtBeginDocument{\let\includegraphics\naturegraphicxincludegraphics}
\makeatletter
\renewenvironment{figure}[1][]{\@float{figure}[#1]}{\end@float}
\renewenvironment{figure*}[1][]{\@dblfloat{figure}[#1]}{\end@dblfloat}
\renewenvironment{table}[1][]{\@float{table}[#1]}{\end@float}
\renewcommand{\subsection}{\@startsection{subsection}{2}{0pt}{12pt}{6pt}{\bfseries}}
\makeatother

\newcommand{\eps}{\varepsilon}
\newcommand{\dd}{\mathrm{d}}
\newcommand{\appendixnumbering}[1]{%
  \setcounter{equation}{0}\renewcommand{\theequation}{#1\arabic{equation}}%
  \setcounter{figure}{0}\renewcommand{\thefigure}{#1\arabic{figure}}%
  \setcounter{table}{0}\renewcommand{\thetable}{#1\arabic{table}}%
  \renewcommand{\theHequation}{#1.\arabic{equation}}%
  \renewcommand{\theHfigure}{#1.\arabic{figure}}%
  \renewcommand{\theHtable}{#1.\arabic{table}}%
}

\begin{document}

\title{Temporal glide symmetry enforces a parity sideband\\selection rule in scalar bulk media}

\author{M. Camacho$^{1}$}

\maketitle

\begin{affiliations}
\item Department of Electronics and Electromagnetism, School of Physics, Universidad de Sevilla, Sevilla, Spain. E-mail: mcamachoa@us.es
\end{affiliations}

\begin{abstract}
Symmetry is a powerful way to control coupling between photonic mode families. In spatially periodic structures, glide symmetry can protect band contacts and suppress stop bands. Here we show a different role for temporal glide, a spatiotemporal counterpart combining reflection with a half-period time translation. In a scalar time-modulated trilayer waveguide, temporal glide imposes an exact selection rule linking frequency conversion to transverse-mode symmetry: the parity content of every Floquet eigenstate alternates with sideband index, up to a state-dependent sign. In scattering, this means that a mode of definite parity can emit only into the opposite transverse parity at odd sidebands and into the same parity at even sidebands. We verify the rule directly in bulk Floquet eigenstates and in finite-section time-domain simulations. An incident odd waveguide mode is converted into an even frequency sideband, while all symmetry-forbidden output channels at the analysed sidebands are suppressed to numerically negligible values. Whereas spatial glide constrains the spectrum, temporal glide in scalar bulk media acts instead as a field-level principle, converting electromagnetic energy between selected modes and frequencies under an exact parity constraint.
\end{abstract}

\section{Temporal glide selects sideband parity}

In periodic electromagnetic structures, spatial glide is a standard route by which symmetry controls spectra through the Bloch phase advance across one unit cell, the electrical length of the cell. A reflection followed by a half-cell translation squares to a full-cell translation: for a Bloch mode with phase advance \(e^{i\beta a}\), the glide channels carry phases \(\pm e^{i\beta a/2}\). These two square-root branches constrain which modes can mix at the Brillouin-zone edge, giving the familiar picture behind glide band sticking, stop-band suppression and dispersion control in electromagnetic structures\cite{crepeau,mittra,hessel,quevedo,dahlberg}, as well as glide-protected topological photonic phases\cite{lu_glide_topological_photonic_crystal}.

Temporal glide is the spatiotemporal operation that combines reflection with a half-period time translation. Photonic time crystals and other time-varying media possess Floquet quasifrequencies, temporal zones, momentum gaps and temporal mode conversion\cite{galiffi_review,asgari_ptc_tutorial,pendry_PTC,engheta_PTC,dikopoltsev_ptc_free_electrons,moussa_temporal_reflection,solis_temporal_mode_conversion}. Time-glide symmetry is also well established in space-time groups, including recent \(2+1\)-dimensional classifications, and in Floquet topological phases\cite{xu_space_time_group,ke_wu_2d_space_time_groups,morimoto_time_glide,peng_time_glide_soti,mochizuki_quantum_walk_time_glide}. If a time-glide spectrum in a scalar bulk medium shows a contact at \(|\Omega T/\pi| = 1\), it is tempting to read it as the temporal analogue of spatial-glide sticking. A folded Floquet diagram, however, cannot settle this. A seam contact in such a plot is ambiguous: it can be a protected degeneracy, or simply two reciprocal Floquet representatives of one state meeting under zone folding, and the dispersion alone does not distinguish the two. We therefore do not decide the question from the spectrum alone. The sharp symmetry statement lies instead in the fields: temporal glide constrains the harmonic structure of the Floquet eigenvectors, as an exact link between transverse parity and sideband index. Here bulk means a relative-permittivity modulation distributed through the guide volume, not a boundary or sheet modulation.

Parity-changing mode conversion by modulation is itself well established. Indirect photonic transitions use a travelling spatiotemporal perturbation whose spatial symmetry is engineered to couple even and odd waveguide modes, transferring both energy and momentum\cite{yu_fan_indirect_transitions,lira_interband_silicon}, and the modulation phase between regions has recently been identified as a symmetry-control parameter for intermodal gap engineering in time-modulated metasurface waveguides\cite{li2026_metasurface}. The result here is different. In the bulk problem the modulation is purely temporal, so the modulation itself supplies no longitudinal wave vector. The constraint is an exact, nonperturbative property of simultaneous Floquet-glide eigenstates at the glide point, beyond a perturbative matrix-element zero of a designed travelling profile. The transverse parity of the harmonic at sideband index \(m\) is fixed to alternate as \((-1)^m\). The familiar perturbative selection rule is therefore the small-modulation face of a stronger harmonic-wise constraint, which holds throughout the ideal scalar-bulk glide protocol.

The distinction is clearest by comparing a synchronous drive, mirror-symmetric at every instant, with a time-glide drive in which the outer layers are shifted by half a period. Each endpoint carries its own exact rule---constant harmonic parity for synchronous switching and alternating harmonic parity for time glide---while a generic phase delay satisfies neither (Fig.~\ref{fig:protocol}). In a globally glide-symmetric finite section, the alternating rule appears as inelastic parity conversion between Floquet sidebands; the same-parity sideband content is then forced to the calibrated numerical floor rather than merely reduced by phase mismatch.

\section{Temporal glide squares to Floquet evolution}

We consider a scalar dielectric waveguide bounded by perfect-electric-conductor plates at \(z=\pm h/2\). The centered trilayer has a middle slab with relative permittivity \(\bar{\eps}_r=1.625\), while the two outer slabs switch between \(\eps_{r,1}=1.25\) and \(\eps_{r,2}=2.0\). The modulation period is \(T\). Let \(U_A\) and \(U_B\) denote the complete evolution maps over the first and second half periods, including temporal interface maps where required, and define the one-period Floquet operator as
\begin{equation}
M=U_B U_A .
\label{eq:floquet_operator}
\end{equation}
For reflection \(P_z:z\mapsto -z\), the time-glide relation gives
\begin{equation}
P_z U_A = U_B P_z .
\label{eq:glide_relation}
\end{equation}
The corresponding temporal-glide operator is
\begin{equation}
G_t=P_z U_A ,
\label{eq:glide_operator}
\end{equation}
and hence
\begin{equation}
G_t^2=P_zU_AP_zU_A=U_B U_A=M .
\label{eq:gt_square}
\end{equation}
For a simultaneous eigenstate of the one-period map and the glide operator,
\begin{equation}
M\psi=\mu\psi,\qquad G_t\psi=g\psi,\qquad \mu=e^{-i\Omega T},
\label{eq:eigenvalues}
\end{equation}
we have \(g^2=\mu\). Because \(G_t^2=M\), the glide eigenvalue is a square root of the Floquet multiplier, \(g=\sigma\sqrt{\mu}\) with \(\sigma=\pm1\). Its magnitude and phase are fixed by \(\mu\); the remaining choice is the binary sign \(\sigma\), the glide-parity label, determined within each Floquet subspace.

This is unlike time-glide Floquet systems with additional internal degrees of freedom\cite{morimoto_time_glide,peng_time_glide_soti,mochizuki_quantum_walk_time_glide}, where the glide operation acts as a matrix in that internal space and can separate the square-root branches into physically distinct sectors. A scalar bulk relative-permittivity modulation has no such internal space, so the two branches \(g=\pm\sqrt{\mu}\) stay tied to the same multiplier.

The more physical question is then what temporal glide controls once the folded bands themselves carry no glide label beyond the Floquet multiplier itself. It controls the eigenmodes. When a Floquet eigenmode is resolved into temporal harmonics, temporal glide forces an alternating transverse parity across the sideband ladder. Write the associated time-dependent Floquet solution as a temporal harmonic expansion,
\begin{equation}
\psi(z,t)=e^{-i\Omega t}\sum_m \phi_m(z)\,e^{-im\Omega_{\rm mod}t},\qquad \Omega_{\rm mod}=2\pi/T .
\label{eq:harmonic_expansion}
\end{equation}
For a glide eigenstate, \(\psi(-z,t+T/2)=g\,\psi(z,t)\), and matching the expansion harmonic by harmonic gives
\begin{equation}
P_z\,\phi_m=\sigma\,(-1)^m\,\phi_m,\qquad
\sigma=g\,e^{i\Omega T/2}=\pm1 ,
\label{eq:selection_rule}
\end{equation}
where \(\sigma\) is real and binary because \(g^2=\mu=e^{-i\Omega T}\). Equation~\eqref{eq:selection_rule} is the scalar-bulk selection rule of temporal glide: each harmonic has a definite transverse parity, and that parity alternates with sideband index. The synchronous drive gives the complementary pattern. There \(P_z\) commutes with the instantaneous evolution, so \(P_z\,\phi_m=\sigma\,\phi_m\) with the same \(\sigma\) for all \(m\): parity is constant across the harmonic ladder. The two endpoints therefore define two exact parity ladders, one constant and one alternating. The Methods section states how gauge choices and degenerate Floquet multipliers are handled.

In scattering terms, the same symmetry statement applies to a globally glide-symmetric finite section fed by a parity-pure incident carrier. The scattering operator then connects sideband \(m\) only to transverse modes of parity \(p\,(-1)^m\). Odd sidebands are parity converters; even sidebands, including the carrier, are parity preservers. This is the operational form tested below.

\begin{figure}[tb]
    \centering
    \includegraphics[width=0.98\textwidth]{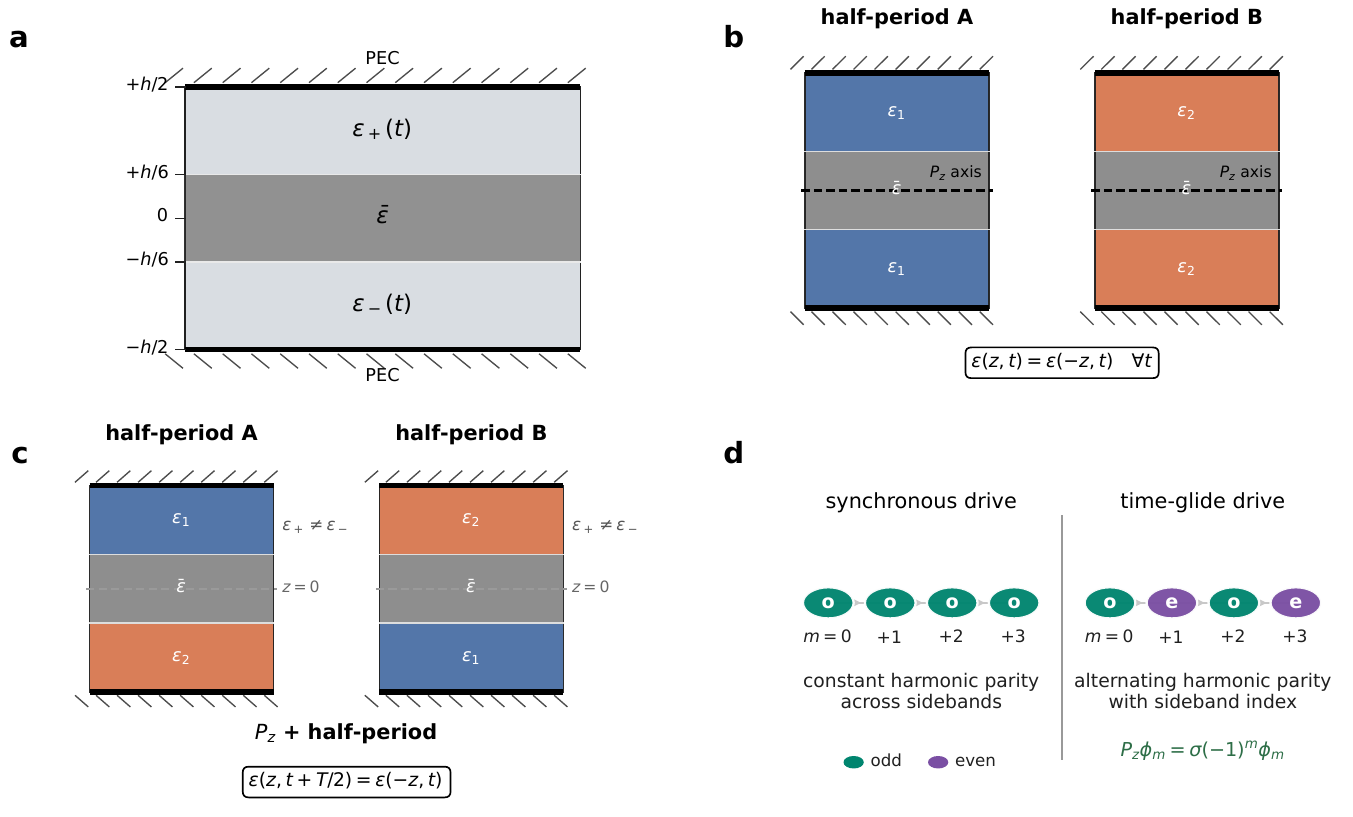}
    \input{figures/fig1_caption.tex}
    \label{fig:protocol}
\end{figure}

Figure~\ref{fig:protocol} turns this rule into the waveguide protocol used below: synchronous modulation keeps every sideband in one transverse-parity family, whereas time glide alternates the allowed family from one sideband to the next.

This algebra also fixes which labels can coexist in a scalar bulk medium. If static reflection and time-glide exchange held pointwise in the same scalar permittivity, \(\eps_r(z,t)=\eps_r(-z,t)\) and \(\eps_r(z,t+T/2)=\eps_r(-z,t)\) would together force \(\eps_r(z,t+T/2)=\eps_r(z,t)\). A nontrivial scalar bulk time-glide drive therefore cannot retain an independent static-parity block decomposition. Static reflection parity is lost for the full drive; the conserved structure is the harmonic-resolved parity pattern of Eq.~\eqref{eq:selection_rule}. The relevant observables are the static-parity hybridization produced during one period and the harmonic-wise alternation that organizes it.

The physical picture is intuitive. In the synchronous protocol, the two outer layers switch together. The modulation is reflection even, so it shifts the even and odd static transverse sectors separately and leaves the coupling between them zero. In the time-glide protocol, the two outer layers switch in opposite order. The reflection-even background remains almost the same, but the drive has acquired a reflection-odd part. That odd part is the useful ingredient: its leading role is to connect even and odd static modes in neighbouring sidebands. A glide spectrum can therefore look close to a static folded dispersion while the fields themselves have been reorganized. The exact statement is the harmonic-wise selection rule derived above.

\section{Time-glide effects in spatiotemporal dispersion diagrams}

The same point appears in the dispersion diagrams once the eigenvectors are inspected. We compute the Floquet spectra with a spatiotemporal Floquet mode-matching formulation in the transverse eigenfunctions of the reference perfect-conductor guide. The analysis shown here is for transverse electric modes, for which temporal jumps conserve \(D_y=\epsilon_0\eps_r E_y\). This operator construction extends to transverse magnetic modes, where the physics is analogous. The calculation is deliberately scalar and bulk, so any parity mixing must come from the time-glide drive itself. The Methods section summarizes the implementation, and Appendices A and B give the basis expansions, jump matrices, guide-mode energy weighting and convergence analysis.

\begin{figure}[tb]
    \centering
    \includegraphics[width=0.98\textwidth]{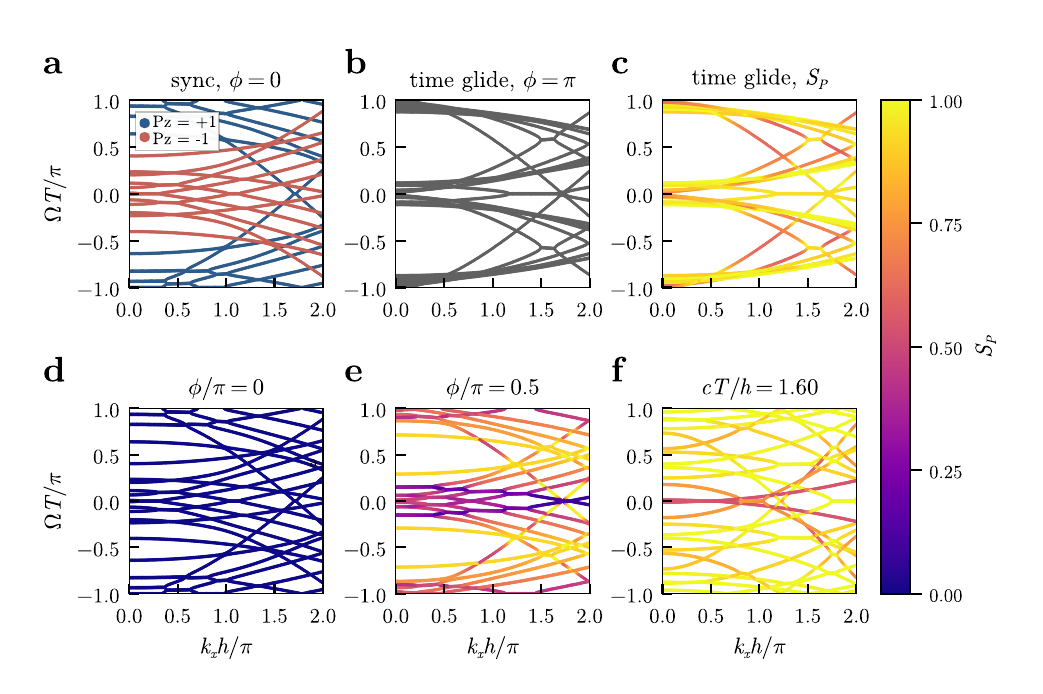}
    \caption{\textbf{Static-parity hybridization from time-glide modulation and its phase and period controls.} \textbf{a,} Floquet spectrum under synchronous modulation (\(\phi=0\)), with branches coloured by exact static reflection parity \(P_z\): blue for \(P_z=+1\) (even sector), coral for \(P_z=-1\) (odd sector). The Floquet operator is block diagonal in static parity and opposite-parity branches may cross without coupling. \textbf{b,} Spectrum under time-glide modulation (\(\phi=\pi\)) at the same period. Reflection exchanges the two half periods instead of leaving each invariant, so static parity is no longer an exact branch label and all branches are shown in grey. \textbf{c,} Same time-glide spectrum coloured by the static-parity mixing entropy \(S_P\) per eigenstate (colormap shared with d--f), where \(S_P=0\) denotes a parity-pure state and \(S_P=1\) equal even/odd energy. The order-unity mixing across the entire spectral window quantifies the reorganization induced by the reflection-odd component of the drive; the median is \(S_P=0.94\). \textbf{d,e,} Phase-delay control. At \(\phi=0\) (d), the same synchronous protocol as in panel a is recoloured by \(S_P\), confirming that the entropy remains below numerical resolution. At intermediate \(\phi=\pi/2\) (e), the reflection-odd component is partly activated and \(S_P\) develops order-unity values. The full glide point \(\phi=\pi\) is shown in panel c. \textbf{f,} Period robustness. Glide spectrum at \(cT/h=1.60\) (compared with \(cT/h=1.27\) in c) shows that changing the modulation period reorganizes the folded dispersion but preserves the order-unity \(S_P\), confirming that the observable is eigenvector hybridization, not a particular branch pattern.}
    \label{fig:spectrum}
\end{figure}

Figure \ref{fig:spectrum} combines the global transverse-electric spectra with the phase and folding controls. In the synchronous drive, static reflection parity is an exact label and opposite-parity branches may cross without coupling (Fig. \ref{fig:spectrum}a). In the time-glide drive, static parity is no longer a good branch label: the same spectral window is reorganized by coupling between the even and odd static sectors (Fig. \ref{fig:spectrum}b,c). If one looks only at the shape of the dispersion, much of the physics is missed. A plotted Floquet branch is a quasifrequency branch, not the trace of a single temporal harmonic. Each point on the curve represents a full periodic eigenvector, and that eigenvector contains a superposition of sidebands. As anticipated by the selection rule above, the symmetry acts on those sideband components, whose transverse parities alternate with the harmonic index.

We quantify this eigenvector change by projecting each Floquet state onto the physical static reflection sectors and evaluating an energy-weighted binary entropy \(S_P\). It is zero for a parity-pure state and unity for equal energy weight in the two sectors. The projection uses the guide-mode energy weighting of the homogeneous reference guide, so the reported weights are physical modal weights in the two reflection sectors.

The sync--glide contrast is already visible in this single number. For the displayed TE spectra we use a centered trilayer with mean permittivity \(\bar{\eps}_r=1.625\), outer-layer values \(\eps_{r,1}=1.25\) and \(\eps_{r,2}=2.0\), contrast \(\Delta\eps_r=0.75\), period \(cT/h=1.27\), basis size \(N=12\) and \(k_xh/\pi\in[0,2]\). The synchronous drive gives \(S_P<10^{-12}\), at numerical zero. The time-glide drive reaches a median \(S_P=0.938\), meaning that typical Floquet eigenvectors carry comparable guided-mode energy in the even and odd sectors.

The hybridization is not unstructured: it follows the exact harmonic pattern of Eq.~\eqref{eq:selection_rule}. To test the rule directly, we propagate each Floquet eigenvector over one period, extract its temporal harmonics \(\phi_m\), and evaluate a violation metric \(\mathcal V\), defined in Methods as the energy fraction carried by the wrong parity sector summed over harmonics and minimized over the binary choice of \(\sigma\). At the glide point the rule holds to machine precision across all computed states, wave numbers, basis sizes, periods and contrasts, with a maximum violation \(\mathcal V=4.7\times10^{-25}\). Away from the glide point the rule fails at order unity, with median violations between \(0.01\) and \(0.45\) and maxima at \(0.5\) for intermediate phases. The alternation is therefore an exact property of the glide drive, not an approximate tendency of strong parity mixing. The direct operator and harmonic checks are given in Methods and Appendices A and B.

\section{Phase-delay and folding controls}

A phase-delay sweep gives a direct way to move between the two exact patterns. We write the upper and lower outer-layer relative permittivities as
\begin{equation}
\eps_{r,+}(t)=\bar{\eps}_r+\frac{\Delta\eps_r}{2}s(t),\qquad
\eps_{r,-}(t)=\bar{\eps}_r+\frac{\Delta\eps_r}{2}s(t+\tau),
\label{eq:phase_delay_results}
\end{equation}
where \(s(t)=\pm1\) and \(\phi=2\pi\tau/T\). At \(\phi=0\), the profile is mirror symmetric at every instant and the harmonic parity is constant in \(m\). At \(\phi=\pi\), reflection exchanges the two half periods and the harmonic parity alternates with \(m\). Between these endpoints, the reflection-odd part of the drive is turned on continuously, and neither exact ladder rule is enforced. The intermediate \(\phi=\pi/2\) spectrum in Fig. \ref{fig:spectrum}e already shows order-unity eigenvector mixing.

This sweep is a controlled interpolation between two symmetry endpoints. The synchronous endpoint has \(S_P\) below numerical resolution and a constant-parity harmonic ladder. The time-glide endpoint has median \(S_P=0.938\) and an exactly alternating ladder. At intermediate phases the violation metric \(\mathcal V\) of the alternating rule reaches order unity. What matters is the change between two exact endpoint patterns, not the monotonicity of every percentile.

Changing the modulation period provides a separate folding control (Fig. \ref{fig:spectrum}f). The period \(T\) changes where the branches appear inside the quasifrequency zone, but it does not remove the field reorganization: for selected time-glide spectra with \(cT/h\) between 1.00 and 1.60, the median \(S_P\) remains between 0.77 and 0.98. This separates the physical signature, static-parity hybridization, from the visual details of Floquet branch folding.

This period control also clarifies why a temporal-zone-edge contact should not be read as a spatial-glide degeneracy. In an extended-zone view, a positive-frequency branch can meet the folded copy \(-\Omega+2\pi/T\). In the scalar temporal problem this can be a crossing of reciprocal Floquet representatives of the same multiplier \(\mu=-1\), not a two-branch sticking event.

\section{Symmetry-selected sideband conversion}

The selection rule can be used as a symmetry-selected sideband-conversion mechanism. In a finite section whose full modulation envelope preserves the same glide relation, a static transverse mode of parity \(p\) can be converted at sideband \(m\) only into transverse modes of parity \(p\,(-1)^m\): the carrier and even sidebands preserve parity, while odd sidebands convert it. A synchronous drive obeys the complementary prediction, with all sidebands in the incident parity sector. The signature is therefore not the mere appearance of even and odd static parities in the output, which a generic phase delay can also produce. It is the sideband-by-sideband assignment itself: each \(m\) must occupy the predicted parity sector and leave the other one empty. In practice, the bulk hybridization statement becomes an output-channel test: project the transmitted field onto the static modes of the unmodulated output guide, resolve them by sideband, and compare the parity content of each \(m\) against the predicted pattern.

We first demonstrate this in an intentionally spectrally isolated converter. The finite section is a two-dimensional transverse-electric (TE) FDTD trilayer embedded between static input and output guides. We write \(\mathrm{TE}_n\) for a TE guide mode with transverse index \(n\). The design target is to convert an incident odd \(\mathrm{TE}_2\) carrier at \(\omega_0 h/c=5.65\) into the glide-selected even \(\mathrm{TE}_3\) sideband at \(m=+1\). The section has length \(L=4h\), \(0.8h\) raised-cosine tapers and sinusoidal outer-layer modulation. The modulation frequency is chosen so that this target channel propagates, while the next higher generated channels are below cutoff. This isolates the useful conversion: the symmetry-selected channel is not buried in a dense multimode ladder.

We define two figures of merit, which answer different questions. The aggregate opposite-parity conversion fraction \(C_{\rm opp}\) is computed from the propagating output power over the analysed sidebands and asks what fraction has changed transverse parity relative to the incident carrier. It tells us whether parity conversion has occurred, but not whether the selection rule is obeyed sideband by sideband. The stricter test is the wrong-parity fraction per sideband, \(W_m\), defined as the fraction of sideband-\(m\) propagating power lying in the parity sector forbidden by Eq.~\eqref{eq:selection_rule}. The glide rule requires \(W_m=0\) for every analysed sideband simultaneously.

\begin{figure}[tb]
    \centering
    \includegraphics[width=0.98\textwidth,height=0.60\textheight,keepaspectratio]{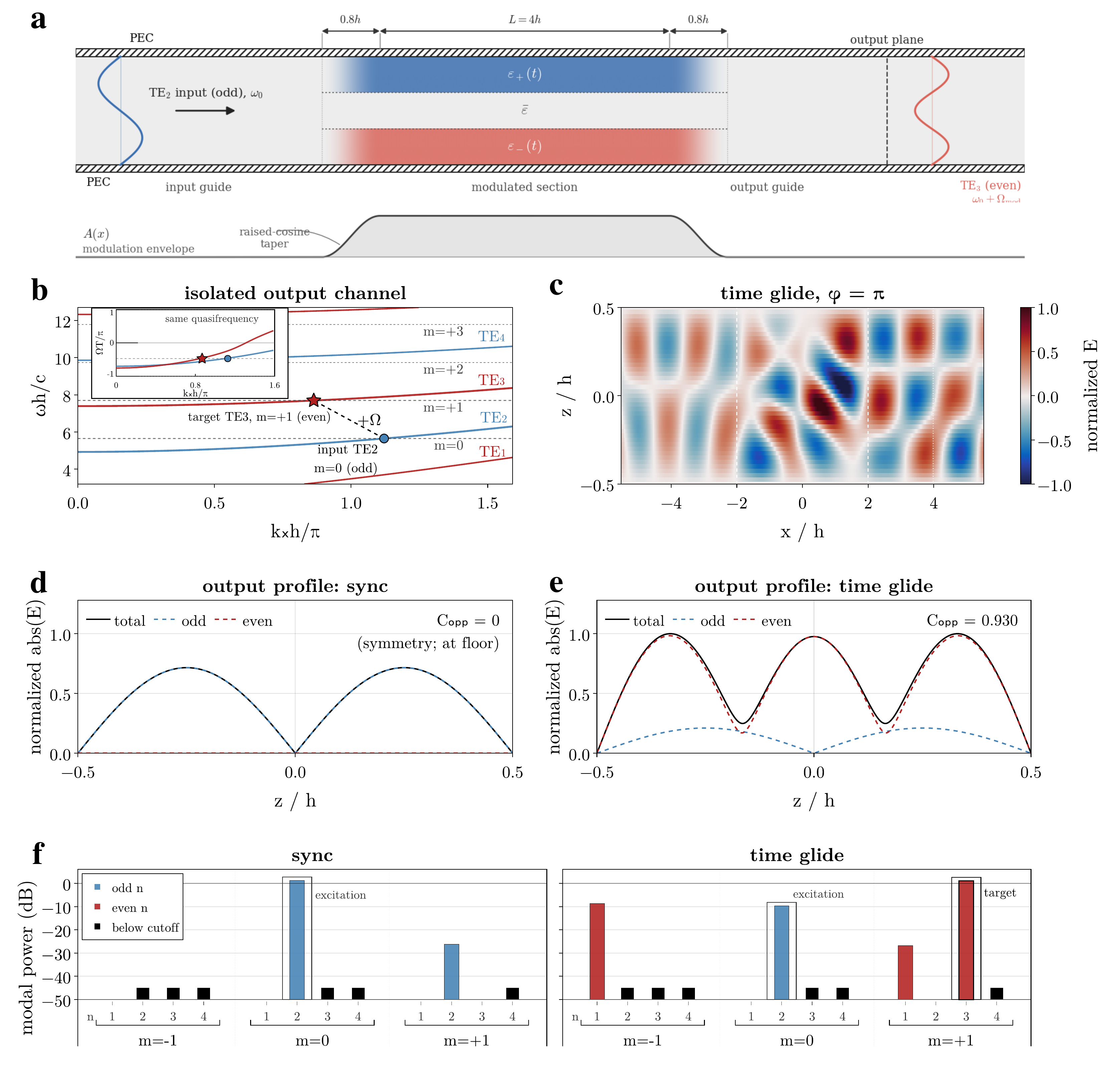}
    \caption{\textbf{Spectrally isolated parity conversion in a time-glide-modulated waveguide section.} \textbf{a,} FDTD geometry for an odd \(\mathrm{TE}_2\) carrier incident on a finite sinusoidally modulated trilayer. \textbf{b,} Output-guide dispersion: \(\mathrm{TE}_2\) at \(m=0\) and \(\mathrm{TE}_3\) at \(m=+1\) propagate, while the next competing generated channels are below cutoff; the inset shows the folded time-glide branches placing input and target at the same quasifrequency. \textbf{c,} Normalized signed instantaneous field for the time-glide drive. \textbf{d,e,} Output-plane modal RMS profiles decomposed into odd and even static-mode projections. The synchronous even projection is zero by symmetry, while time glide gives \(C_{\rm opp}=0.930\). \textbf{f,} Modal-resolved output spectra for \(m=-1,0,+1\), normalized to the incident \(\mathrm{TE}_2\) carrier. The time-glide response is dominated by the target \(\mathrm{TE}_3\), \(m=+1\) even sideband, which contains \(90.1\%\) of the generated propagating modal power; forbidden sideband parities sit at the numerical-error floor.}
    \label{fig:fdtd-scattering}
\end{figure}

Figure~\ref{fig:fdtd-scattering} summarizes the finite-section geometry, output-guide dispersion, FDTD profiles and sideband-resolved powers for both drives. In the synchronous case, all sidebands remain in the incident odd sector; the even sector is symmetry-forbidden and below numerical error. In the time-glide case, the output has the three-lobe structure of the even \(\mathrm{TE}_3\) channel at \(\omega_0+\Omega_{\rm mod}\). The sideband-resolved powers give \(C_{\rm opp}=0.930\). About \(90\%\) of the generated propagating power lies in the target \(\mathrm{TE}_3\), \(m=+1\) sideband, and every analysed \(W_m\) is below the numerical-error floor.

\begin{figure}[tb]
    \centering
    \includegraphics[width=0.98\textwidth]{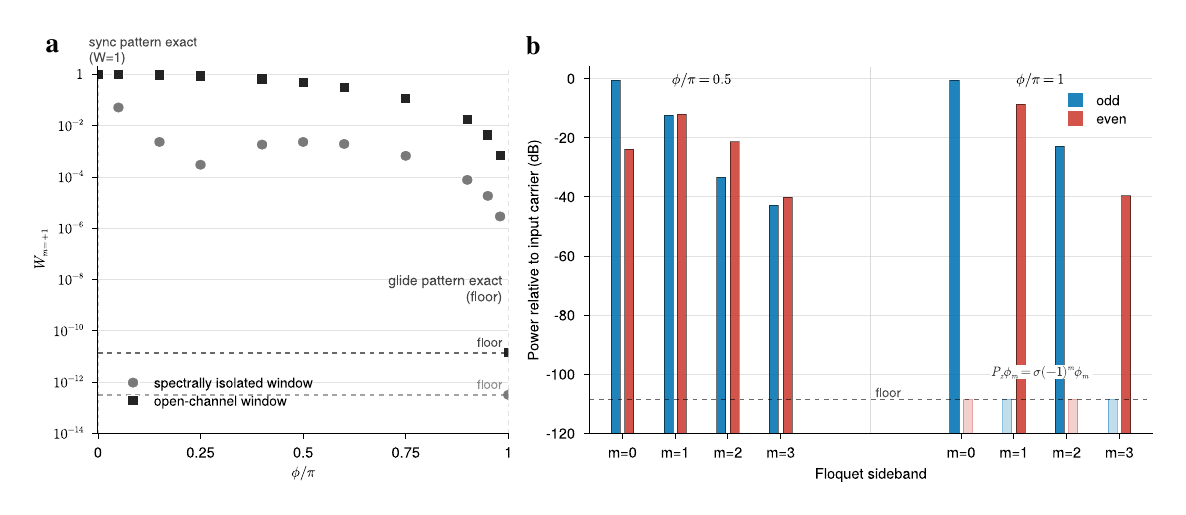}
    \caption{\textbf{Exact parity--sideband selection rule in finite-section scattering.} \textbf{a,} Wrong-parity fraction \(W_{m=+1}\) versus modulation phase delay \(\phi\) (logarithmic scale), where \(W_m\) is the fraction of sideband-\(m\) power in the parity sector forbidden by the glide alternation rule. Data are shown for the spectrally isolated window of Fig.~\ref{fig:fdtd-scattering} and for the configuration at \(\omega_0h/c=8.0\), \(cT/h=2.0\), in which the forbidden odd channel \(\mathrm{TE}_4\) at \(m=+1\) is propagating with only residual propagation-constant mismatch. The sweep interpolates between the two exact patterns: the synchronous rule at \(\phi=0\), for which \(W_{m=+1}=1\) only relative to the glide-rule convention, and the glide rule at \(\phi=\pi\), where \(W_{m=+1}\) reaches the numerical-error floor. These two endpoints are the parity-preserving (\(\phi=0\)) and parity-converting (\(\phi=\pi\)) modes of the switchable converter discussed in the main text. Intermediate-phase values are controlled by phase matching and are configuration-specific. In the open-channel case \(W_{m=+1}\) remains \(6.9\times10^{-4}\) at \(\phi/\pi=0.98\), many orders above the floor, before collapsing at the exact glide point. \textbf{b,} Sideband-resolved parity content at \(\phi=\pi/2\) (left) and \(\phi=\pi\) (right) for the open-channel configuration, for \(m=0,\dots,+3\). At the glide point the parity alternates exactly with \(m\) as required by Eq.~\eqref{eq:selection_rule}: odd, even, odd, even, with wrong-parity fractions at the numerical-error floor (faded), while at \(\phi=\pi/2\) both parities populate each sideband. The dashed line marks the calibrated floor of each run.}
    \label{fig:selection-rule}
\end{figure}

To check that the parity selection rule is symmetry enforced, and not accidental mixing in this geometry, we repeat the finite-section calculation while varying the phase delay \(\phi\) between the two modulated outer layers. Figure~\ref{fig:selection-rule}a shows the resulting \(W_{m=+1}\) on a logarithmic scale from the synchronous endpoint to the glide endpoint, for both the isolated converter of Fig.~\ref{fig:fdtd-scattering} and a complementary open-channel experiment at \(\omega_0h/c=8.0\) and \(cT/h=2.0\), where the same-parity odd \(\mathrm{TE}_4\) channel at \(m=+1\) is propagating. The two numerical experiments probe complementary aspects of the physics. In the converter design, unwanted channels are deliberately suppressed so that the target sideband is dominant. In the open-channel experiment, more modes are allowed to propagate, and the same forbidden sector is readily populated at intermediate phase. In both cases, however, the wrong-parity content stays finite until the exact glide point and then collapses to the numerical-error floor. Phase mismatch can suppress a channel; time-glide symmetry forbids it.

The open-channel experiment makes this second test explicit. Figure~\ref{fig:selection-rule}b resolves the same run by sideband and parity. Alongside the same-parity \(\mathrm{TE}_4\) channel, the even \(\mathrm{TE}_3\) channel and additional channels at \(m=+2,+3\) are also open. At intermediate phase the forbidden channel carries a large fraction of the \(m=+1\) sideband power, and both parities populate the sideband ladder. At the glide point, with the same channels open, the ladder collapses onto the predicted sequence for an odd incident carrier: odd at \(m=0\), even at \(m=+1\), odd at \(m=+2\) and even at \(m=+3\), with the forbidden bars at the numerical-error floor. Thus the glide zero is neither a cutoff artefact nor a weak-channel accident.

\FloatBarrier

\section{Experimental feasibility}

The experimentally robust observable is the sideband-resolved excitation of static transverse-parity sectors. A microwave test can use the same guide, modulation depth and modulation frequency, and change only the relative phase between the two outer-region drives. A near-field scan at the output, demodulated at \(f_0+m f_{\rm mod}\), would be projected onto the static modes of the unmodulated guide and grouped by parity. The synchronous run is the equal-amplitude frequency-conversion reference and calibrates residual asymmetry; the time-glide run should switch the allowed static parity at odd sidebands while preserving it at even sidebands. Repeating the measurement with the opposite incident parity would test the same ladder from the other side.

The large contrast used in Fig.~\ref{fig:fdtd-scattering} makes the converted sideband visible in a short section; it is not a fundamental requirement. Reducing the effective swing from \(\Delta\eps_r=0.65\) to \(0.05{-}0.10\) lowers the generated sideband powers by about \(16{-}22\) dB under the perturbative scaling, but does not change the parity ratio \(W_m\). The critical tolerance is instead residual breaking of the glide relation. In the open-channel sweep, the near-glide points obey \(W_{+1}\simeq1.7(1-\phi/\pi)^2\); keeping the relative phase within \(\Delta\phi\simeq4^\circ\) of the glide point therefore keeps \(W_{+1}<10^{-3}\), below the phase-matching residual of the spectrally isolated converter. At \(f_{\rm mod}=3.25\,{\rm GHz}\), this is a timing precision of about \(3.4\,{\rm ps}\), well within standard phase-locked drive synthesis. Glide-symmetric loss or tapering reduces signal-to-noise but does not populate the forbidden parity sector; only differential loss, phase error or drive imbalance enters this symmetry-breaking budget. Parallel-plate guides with phase-locked tunable inclusions, or transmission-line/metasurface-network implementations, provide microwave routes to this scalar TE limit; Appendix D gives the corresponding scales and controls.

\section{Discussion and outlook}

Time-modulated waveguide spectra should be read through their fields, not only through their folded branches. A glide spectrum that resembles a static folded spectrum is not evidence that the glide drive is weak, and an apparent temporal seam crossing is not by itself evidence of spatial-glide-like sticking. The observables that survive this ambiguity are modal: \(S_P\) for bulk parity mixing, \(\mathcal V\) for the harmonic rule, and \(W_m\) for the wrong-parity power in a measured sideband. They test the fields directly, without having to decide which folded branch connects to which.

The finite-section simulation translates this into a scattering statement. The measured contrast is the sideband-resolved parity content. The hierarchy is quantitative: phase matching suppresses the forbidden channel to the \(10^{-3}\) level in the engineered window, while glide symmetry places it at the numerical floor even when the channel is open and visibly populated at intermediate phase, more than seven orders of magnitude below the phase-matching residual. The phase delay itself is a diagnostic rather than an optimization parameter: it separates dispersion shifts from the reflection-even background and parity recombination from the reflection-odd drive.

In microwave terms, the result is a selection rule for a multichannel frequency converter. The output channels are labelled by a transverse guide mode and a sideband index \((n,m)\). Temporal glide does not merely change how strongly those channels are driven; it removes half of them from the conversion matrix at the symmetry point. For an odd incident carrier, the allowed output ladder is odd at \(m=0\), even at \(m=\pm1\), odd at \(m=\pm2\), and so on. More generally, a protocol that combines an order-\(N\) transverse operation with a time shift \(T/N\) would organize the sideband ladder by \(m\bmod N\) (for example, an \(N\)-fold rotation in a guide of suitable cross-section). Viewed in the synthetic frequency dimension spanned by the sideband index\cite{yuan_synthetic_dimension_photonics}, the alternating parity ladder is a nonsymmorphic sublattice structure set by protocol timing, with no need to prescribe individual couplings. For design, the timing of the modulation banks sets the channel pattern, without fine tuning each modal coupling separately.

A single device therefore interpolates between two complementary selection rules through the relative phase \(\phi\) alone. At \(\phi=0\) the drive is reflection-even and parity is preserved across the whole sideband ladder; at \(\phi=\pi\) it acquires the reflection-odd component and parity alternates with \(m\). The synchronous endpoint thus realizes a parity-preserving converter and the time-glide endpoint a parity-converting one, on the same guide and modulation hardware, with the forbidden sector held below the calibrated numerical floor at each endpoint.

The scalar-bulk assumptions delimit the claim, but they also indicate the next microwave designs to try. Modulated sheets and metasurfaces\cite{wang_metasurface_ptc,li2026_metasurface,lyubarov_cluster_states}, switched boundaries\cite{solis_temporal_mode_conversion}, acoustic or elastic implementations\cite{zanjani_phonon_isolation,chen_acoustic_mode_transitions,chen_elastic_temporal_interfaces,samak_phononic_temporal_interfaces}, vectorial or bianisotropic media, active systems and network models can all carry additional ports or internal degrees of freedom. To obtain independent time-glide sectors of the type discussed in Floquet classifications\cite{morimoto_time_glide,peng_time_glide_soti,mochizuki_quantum_walk_time_glide}, the glide operation must act on such an internal channel---for example two coupled transmission lines exchanged by mirror symmetry, two coupled guides exchanged by \(P_z\), two polarizations, or bianisotropic field components. In scalar relative-permittivity modulation, the direct bulk signature is instead an exact parity--sideband alternation rule, turning the guide into a frequency converter whose parity selectivity is enforced rather than engineered.

\section{Methods}

\subsection{Trilayer geometry and drive protocols}

The model waveguide is bounded by perfect-electric-conductor plates at \(z=\pm h/2\). The dielectric is a centered trilayer: the middle third of the guide is held at \(\bar{\eps}_r=1.625\), while the two outer thirds are temporally switched. Lengths are reported in units of the guide height \(h\), times in units of \(h/c\), frequencies in units of \(c/h\), and longitudinal wave numbers as \(k_xh\). Unless otherwise stated, the bulk Floquet calculations use outer-layer values \(\eps_{r,1}=1.25\) and \(\eps_{r,2}=2.0\), contrast \(\Delta\eps_r=0.75\), dimensionless period \(cT/h=1.27\), and 12 transverse basis functions.

The synchronous protocol switches the upper and lower outer layers in phase, so each instantaneous profile is mirror symmetric. The time-glide protocol switches them in opposite order: reflecting the first half-period profile gives the second half-period profile. The central layer is unchanged in both protocols. In this centered geometry, the synchronous drive changes the reflection-even part of the scalar operator, while the time-glide drive reverses the reflection-odd part between the two half periods. The bulk mode-matching calculations use piecewise-static switching, whereas the finite-section FDTD validation uses a sinusoidal modulation to isolate the first temporal sidebands. These are not compared as identical waveforms; they are two realizations of the same \(\phi=0\) and \(\phi=\pi\) symmetry relations, and the selection rule depends on that symmetry, not on the waveform shape.

\subsection{Bulk Floquet mode matching}

We compute the main-text spectra with spatiotemporal Floquet mode matching for TE polarization. The tangential electric field is expanded in the sine eigenfunctions of the PEC guide. Each half period is piecewise static, so the modal amplitudes are propagated through the half-period evolution and matched across temporal dielectric jumps by conserving displacement rather than electric field. This is the temporal-interface condition for scalar permittivity modulation.

The one-period Floquet operator is the ordered product of static evolutions and displacement-conserving jumps: two intervals at the symmetry endpoints and, for a generic phase delay, four intervals with jumps at \(t=\tau,T/2,T/2+\tau\) and \(T\). We set \(\Omega T=-\arg\mu\in(-\pi,\pi]\) and store \(\log|\mu|\) separately. We keep states with \(|\mu|\neq1\) in the statistics, so reciprocal growing and decaying partners enter \(S_P\) and \(\mathcal V\) on the same footing as stable states.

\subsection{Static-parity modal analysis}

To measure the static-parity content, each Floquet eigenvector is projected onto the even and odd eigenspaces of the physical reflection \(P_z\). The sector weights \(w_\pm\), with \(w_++w_-=1\), use the guide-mode energy weighting of the homogeneous reference guide at the mean permittivity \(\bar{\eps}_r=1.625\). The reported mixing is therefore a physical modal weight in the two reflection sectors. The entropy \(S_P=-\sum_{\pm}w_\pm\log_2 w_\pm\) is zero for a state lying entirely in one static parity sector and is one for equal energy in the two sectors.

Appendix B verifies that the parity weights are unchanged when the same guide-mode energy weighting is evaluated either in field form or in the modal basis. The sectors are fixed by the physical reflection operator and weighted with the same normalization used for the modal amplitudes\cite{joannopoulos_photonic_crystals}.

\subsection{Harmonic-resolved selection-rule verification}

We test the selection rule of Eq.~\eqref{eq:selection_rule} directly on the Floquet eigenstates. Each eigenvector is propagated over one period with the analytic piecewise-static evolution, sampled at 256 points, multiplied by \(e^{+i\Omega t}\) and Fourier transformed to obtain \(\phi_m\) for \(|m|\le8\). The quasifrequency gauge is fixed to the first Floquet zone. Shifting \(\Omega\) by \(\Omega_{\rm mod}\) only relabels \(m\) by one step and exchanges the two allowed parity ladders, so the alternation rule is gauge invariant. Each harmonic is projected onto the even and odd sectors with the same guide-mode energy weighting used for \(S_P\). The violation metric is
\begin{equation}
\mathcal V=\min_{\sigma=\pm1}\ \frac{\sum_m \mathcal U_m^{\,\mathrm{wrong}(\sigma)}}{\sum_m \mathcal U_m^{\,\mathrm{total}}},
\label{eq:violation_metric}
\end{equation}
where \(\mathcal U_m\) is the guided-mode energy weight of harmonic \(m\). The minimization over \(\sigma\) is only a binary choice between the two glide-allowed ladders for that eigenstate: even at \(m=0\), odd at \(m=1\), even at \(m=2\) for \(\sigma=+1\), or the opposite sequence for \(\sigma=-1\). After this choice, \(\mathrm{wrong}(\sigma)\) is the parity sector forbidden by that ladder at each harmonic. Degenerate Floquet multipliers are handled by diagonalizing \(G_t\) within the degenerate subspace before evaluating \(\mathcal V\). At the glide point, across all \(k_x\), basis sizes \(N=12\) and \(20\), periods \(cT/h\in\{1.27,1.60\}\) and contrasts \(\Delta\eps_r\in\{0.2,0.75\}\), the maximum violation is \(4.7\times10^{-25}\) and the worst-case median is \(1.8\times10^{-28}\). At phases \(\phi/\pi\in\{0.25,0.5,0.75\}\) the median violation ranges from \(0.010\) to \(0.445\) with maxima at \(0.5\). Independent operator checks, evaluated with the Frobenius norm, give relative residuals \(\|P_zU_A-U_BP_z\|/\|U_BP_z\|\le4.4\times10^{-14}\), \(|g^2-\mu|\le7.7\times10^{-14}\) and \(|\,|\sigma|-1\,|\le3.9\times10^{-14}\), with the sign of \(\sigma\) matching the harmonic parity pattern in every computed state.

\subsection{Control sweeps and convergence}

The phase-delay sweep independently modulates the two outer layers with a relative delay \(\tau\), reported as \(\phi=2\pi\tau/T\). The endpoint \(\phi=0\) is synchronous switching, and \(\phi=\pi\) is time-glide exchange. This sweep is used as a control knob because it continuously turns on the reflection-odd drive at fixed contrast.

We also vary modulation period, contrast, modulation frequency, basis size and reference guide-mode energy weighting. Increasing the TE basis from \(N=4\) to \(N=20\) leaves the synchronous drive parity-pure below numerical resolution and keeps the time-glide drive strongly hybridized. For \(N=12\), the time-glide value is the main-text median \(S_P=0.938\). Recomputing the TE entropy with the homogeneous, first-half-period and second-half-period reference mass matrices gives agreement to six significant digits. Period, contrast and frequency sweeps are summarized in Appendix B.

\subsection{Finite-section FDTD scattering simulation}

The finite-section calculation in Fig.~\ref{fig:fdtd-scattering} uses a two-dimensional TE FDTD solver for fields \((E_y,H_x,H_z)\). PEC plates impose \(E_y=0\) at the transverse boundaries, and third-order polynomial absorbing layers of thickness \(1.2h\) terminate the longitudinal ends. These layers suppress reflections in the finite computational box, and the reported quantities are symmetry-resolved modal power fractions at the output.

The time-dependent dielectric is updated in displacement form. The scheme advances \(D_y\) from the curl of \(H\), evaluates \(\eps_r(x,z,t)\), and then reconstructs \(E_y=D_y/(\epsilon_0\eps_r)\). This keeps \(D_y\) continuous through the temporal modulation. The Fig.~\ref{fig:fdtd-scattering} case uses \(\bar{\eps}_r=1.625\), outer-layer values 1.3 and 1.95, contrast 0.65, grid steps \(\Delta x=0.04h\) and \(\Delta z=0.025h\), and Courant factor 0.55. The modulated section is \(4h\) long with \(0.8h\) raised-cosine tapers on both sides. The source is a current sheet with the odd \(\mathrm{TE}_2\) transverse profile, driven at \(\omega_0h/c=5.65\) with a \(\sin^2\) turn-on over 12 modulation periods. The sinusoidal modulation period is \(cT/h=3.07486\). At \(h=30\,{\rm mm}\), the carrier, modulation and target-sideband frequencies are 8.99, 3.25 and 12.24 GHz, respectively.

The first configuration isolates one generated opposite-parity channel. The target \(\mathrm{TE}_3\), \(m=+1\) sideband is just above cutoff, whereas the next higher generated channels are below cutoff. The output field is recorded in the static guide after the section, projected onto the static sine modes of the homogeneous reference guide, and demodulated by a Hann-windowed Fourier component at \(\omega_0+m\Omega_{\rm mod}\). The simulations run for 72 modulation periods, with a 40-period steady-state record beginning after 32 periods. For a propagating TE mode with modal amplitude \(a_{n,m}\) and longitudinal propagation constant \(\beta_n(\omega_m)\), the modal power is proportional to \(\beta_n(\omega_m)|a_{n,m}|^2/(\omega_m\mu_0)\), up to the common transverse normalization that cancels in the reported fractions; modes with imaginary \(\beta_n\) are treated as evanescent and excluded from the propagating-power sums. The opposite-parity conversion fraction \(C_{\rm opp}\) is the fraction of propagating output modal power lying in transverse modes of opposite parity to the incident mode. The wrong-parity fraction per sideband is \(W_m=P_m^{\rm wrong}/(P_m^{\rm wrong}+P_m^{\rm right})\), with the right parity at sideband \(m\) given by Eq.~\eqref{eq:selection_rule}. The numerical floor of \(W_m\) is calibrated from the even-sector content of the synchronous control and a grid-refined glide run, giving \(3.2\times10^{-13}\) for the first configuration and \(1.4\times10^{-11}\) for the second. At \(m=-1\) the odd static sector is below cutoff in the first configuration, so \(W_{-1}=0\) there is cutoff-limited rather than a symmetry diagnostic. Appendix C gives the sideband set, spectral window and convergence checks.

The second configuration uses \(\omega_0h/c=8.0\) and \(cT/h=2.0\) with the same grid policy, tapers, time window and source ramp. At these parameters the same-parity odd channel \(\mathrm{TE}_4\) at \(m=+1\) is propagating in the output guide with residual wavenumber mismatch \(|\Delta k|L=5.7\); it is nevertheless demonstrably populated at intermediate phase, with \(W_{m=+1}=0.48\) at \(\phi=\pi/2\). Additional channels are open at \(m=+2,+3\). The analysed sideband set is \(m=-1,\dots,+3\). The near-glide scaling quoted in the feasibility discussion is a least-squares fit of \(W_{+1}\) to the open-channel sweep points with \(\phi/\pi\ge0.90\). Grid, time-window, output-plane, sideband-range and source-ramp controls preserve the same hierarchy in both configurations: the synchronous and glide drives realize their respective exact parity patterns at the calibrated floors, while intermediate phases violate both.

As an independent check, a COMSOL transient simulation of the same scalar TE geometry reproduces the selection-rule hierarchy. In the target \(\mathrm{TE}_2\)-to-\(\mathrm{TE}_3\), \(m=+1\) channel, the synchronous signal is numerically negligible in the raw out-of-plane field-intensity ratio, which is used only as a hierarchy check rather than as a calibrated floor. The glide run places about \(89\%\) of the generated modal power in that target sideband. Solver and export details are given in Appendix C.

\section*{Data availability statement}

The numerical data supporting the figures and metrics are available from the corresponding author upon reasonable request.

\section*{Code availability statement}

The code used to generate the Floquet spectra, hybridization metrics, phase-delay sweeps, finite-section FDTD scattering simulations and convergence checks is available from the corresponding author upon reasonable request.

\section*{Acknowledgments}

This work was supported by Grant PID2023-148281NB-I00 funded by MICIU\slash AEI\slash 10.13039\slash 501100011033 and by the European Regional Development Funds/European Union (ERDF/EU).

\section*{Author contributions}

M.C. conceived the project, developed the formalism and numerical methods, performed the Floquet mode-matching and FDTD simulations, designed the figures, and wrote the manuscript.

\section*{Competing interests statement}

The author declares no competing interests.

\section*{References}

\input{2026_Time-Glide_arxiv.bbl}
\clearpage
\section*{Appendix A---Floquet mode matching and glide algebra}
\appendixnumbering{A}

The TE mode-matching calculation expands the tangential electric field in the PEC sine basis,
\begin{equation}
E_y(x,z,t)=e^{ik_xx}\sum_{n=1}^{N}a_n(t)\varphi_n(z),\qquad
\varphi_n(z)=\sqrt{\frac{2}{h}}\sin\frac{n\pi(z+h/2)}{h}.
\label{eq:app_te_expansion}
\end{equation}
Within a frozen material profile \(s\), Galerkin projection gives
\begin{equation}
\mathbf M_{\eps}^{(s)}\ddot{\bm a}=-c^2\mathbf S\bm a,\qquad
\left[\mathbf M_{\eps}^{(s)}\right]_{mn}=\int_{-h/2}^{h/2}\eps_s(z)\varphi_m(z)\varphi_n(z)\,\dd z,
\label{eq:app_mass}
\end{equation}
with
\begin{equation}
\mathbf S_{mn}=\left(k_x^2+\frac{n^2\pi^2}{h^2}\right)\delta_{mn}.
\label{eq:app_stiffness}
\end{equation}
The propagated state is \(q=(\bm a,\dot{\bm a})^T\). During the same frozen interval,
\begin{equation}
\dot q=\mathcal A_s q,\qquad
\mathcal A_s=
\begin{pmatrix}
0&I\\
-c^2\left(\mathbf M_{\eps}^{(s)}\right)^{-1}\mathbf S&0
\end{pmatrix},
\qquad
\mathbf F_s(\Delta t)=\exp(\mathcal A_s\Delta t).
\label{eq:app_static_evolution}
\end{equation}
The matrix \(\mathbf M_{\eps}^{(s)}\) is assembled from analytic layer integrals over the lower, central and upper thirds of the guide.

A temporal dielectric jump is imposed by continuity of the projected displacement \(D_y=\epsilon_0\eps_r E_y\). In this second-order state convention, the jump from profile \(s^-\) to \(s^+\) is
\begin{equation}
q^+=\mathbf J_{s^+\leftarrow s^-}q^-,\qquad
\mathbf J_{s^+\leftarrow s^-}=
\begin{pmatrix}
\left(\mathbf M_{\eps}^{(s^+)}\right)^{-1}\mathbf M_{\eps}^{(s^-)}&0\\
0&\left(\mathbf M_{\eps}^{(s^+)}\right)^{-1}\mathbf M_{\eps}^{(s^-)}
\end{pmatrix}.
\label{eq:app_jump}
\end{equation}
The two endpoint protocols then read, in the convention used here,
\begin{equation}
U_A=\mathbf J_{B\leftarrow A}\mathbf F_A(T/2),\qquad
U_B=\mathbf J_{A\leftarrow B}\mathbf F_B(T/2),\qquad
M=U_BU_A .
\label{eq:app_uab}
\end{equation}
For a generic phase delay \(0<\tau<T/2\), the period contains four frozen intervals and displacement-conserving jumps at \(t=\tau,T/2,T/2+\tau\) and \(T\). If the corresponding profiles are \(s_1,\ldots,s_4\), the one-period map is
\begin{equation}
M_\phi=\mathbf J_{s_1\leftarrow s_4}\mathbf F_{s_4}(T/2-\tau)
\mathbf J_{s_4\leftarrow s_3}\mathbf F_{s_3}(\tau)
\mathbf J_{s_3\leftarrow s_2}\mathbf F_{s_2}(T/2-\tau)
\mathbf J_{s_2\leftarrow s_1}\mathbf F_{s_1}(\tau).
\label{eq:app_four_interval}
\end{equation}

The reflection matrix in the sine basis is diagonal:
\begin{equation}
\Pi_{mn}=(-1)^{n+1}\delta_{mn},\qquad
P_z=
\begin{pmatrix}
\Pi&0\\
0&\Pi
\end{pmatrix}.
\label{eq:app_reflection_matrix}
\end{equation}
At the glide point the two half-period maps are exchanged by reflection, \(P_zU_A=U_BP_z\), and \(G_t=P_zU_A\) therefore obeys \(G_t^2=M\). The finite-basis operator audit gives
\begin{equation}
\frac{\|P_zU_A-U_BP_z\|_{\rm F}}{\|U_BP_z\|_{\rm F}}\le 4.4\times10^{-14},
\qquad
|g^2-\mu|\le7.7\times10^{-14},
\label{eq:app_operator_audit}
\end{equation}
with the sign of \(\sigma=g e^{i\Omega T/2}\) matching the harmonic parity pattern for every computed glide eigenstate.

\section*{Appendix B---Guide-mode energy weighting and parity analysis}
\appendixnumbering{B}

The static reflection projectors are \(P_\pm=(I\pm P_z)/2\). The parity weights are not coefficient counts; they use the guide-mode energy weighting of the homogeneous reference guide at \(\bar\eps_r=1.625\),
\begin{equation}
\langle q_1,q_2\rangle_{\rm g}=q_1^\dagger \mathbf Q(k_x)q_2,\qquad
\mathbf Q_{\rm TE}(k_x)=
\begin{pmatrix}
\mathbf S(k_x)&0\\
0&\mathbf M_{\bar\eps}/c^2
\end{pmatrix}.
\label{eq:app_energy_metric}
\end{equation}
For a Floquet eigenvector \(v\),
\begin{equation}
w_\pm=\frac{\langle P_\pm v,P_\pm v\rangle_{\rm g}}
{\langle P_+v,P_+v\rangle_{\rm g}+\langle P_-v,P_-v\rangle_{\rm g}},
\qquad
S_P=-\sum_{\pm}w_\pm\log_2w_\pm .
\label{eq:app_sp}
\end{equation}
Thus \(S_P=0\) denotes a parity-pure state, while \(S_P=1\) means equal guided-mode energy in the two static parity sectors.

The same weights have a direct field interpretation. For TE polarization, Eq.~\eqref{eq:app_energy_metric} is the Galerkin form of
\begin{equation}
U_{\rm TE}=\frac{1}{2}\int_{-h/2}^{h/2}\left[
\epsilon_0\bar\eps_r |E_y(z)|^2+
\mu_0\left(|H_x(z)|^2+|H_z(z)|^2\right)\right]\,\dd z .
\label{eq:app_field_energy}
\end{equation}
Projecting with \(P_\pm\) before reconstructing the field gives the same result as projecting the modal vector, because the sine basis is ordered by the physical reflection parity. Recomputing the displayed TE data in both forms gives the same \(S_P\) values to the printed precision.

For the harmonic selection-rule check, each propagated eigenstate is Fourier analysed into \(\phi_m\), and each harmonic is projected with the same guide-mode weighting. The violation metric is
\begin{equation}
\mathcal V=\min_{\sigma=\pm1}
\frac{\sum_m\mathcal U_m^{{\rm wrong}(\sigma)}}{\sum_m\mathcal U_m^{\rm total}},
\label{eq:app_v_metric}
\end{equation}
where the minimization chooses which of the two glide-allowed parity ladders belongs to that eigenstate. Degenerate Floquet multipliers are treated by diagonalizing \(G_t\) in the degenerate subspace before evaluating \(\mathcal V\).

In the modulation-frequency rows of Table~\ref{tab:app_bulk_controls}, \(\omega_{c,1}\) denotes the cutoff angular frequency of the \(\mathrm{TE}_1\) mode in the homogeneous reference guide. The rows collect independent sweeps, so nominally identical parameter points can differ in the last quoted digit because the sampled \(k_x\) windows are not exactly the same.

\begin{table}[tb]
\centering
\caption{Convergence and control summary for the TE bulk calculation. Values below \(10^{-12}\) are at numerical zero for the plotted statistics.}
\label{tab:app_bulk_controls}
\small
\begin{tabular}{llc}
\toprule
Sweep & value & median \(S_P\)\\
\midrule
basis \(N\) & 4  & 0.492\\
basis \(N\) & 8  & 0.717\\
basis \(N\) & 12 & 0.939\\
basis \(N\) & 16 & 0.961\\
basis \(N\) & 20 & 0.957\\
period \(cT/h\) & 1.00 & 0.773\\
period \(cT/h\) & 1.27 & 0.938\\
period \(cT/h\) & 1.60 & 0.980\\
contrast \(\Delta\eps_r\) & 0 & \(<10^{-12}\)\\
contrast \(\Delta\eps_r\) & 0.15 & 0.127\\
contrast \(\Delta\eps_r\) & 0.30 & 0.362\\
contrast \(\Delta\eps_r\) & 0.50 & 0.716\\
contrast \(\Delta\eps_r\) & 0.75 & 0.939\\
modulation \(\omega_m/\omega_{c,1}\) & 0.5 & 0.305\\
modulation \(\omega_m/\omega_{c,1}\) & 1.0 & 0.948\\
modulation \(\omega_m/\omega_{c,1}\) & 2.0 & 0.935\\
modulation \(\omega_m/\omega_{c,1}\) & 4.0 & 0.560\\
\bottomrule
\end{tabular}
\end{table}

\begin{table}[tb]
\centering
\caption{Harmonic-rule violation statistics over the tested \(k_x\) grids, periods, contrasts and TE basis sizes.}
\label{tab:app_v_stats}
\small
\begin{tabular}{cccc}
\toprule
\(\phi/\pi\) & median \(\mathcal V\) & p90 \(\mathcal V\) & max \(\mathcal V\)\\
\midrule
0.25 & \(4.38\times10^{-1}\) & \(4.92\times10^{-1}\) & \(5.00\times10^{-1}\)\\
0.50 & \(4.45\times10^{-1}\) & \(4.93\times10^{-1}\) & \(5.00\times10^{-1}\)\\
0.75 & \(4.41\times10^{-1}\) & \(4.91\times10^{-1}\) & \(5.00\times10^{-1}\)\\
1.00 & \(1.83\times10^{-28}\) & \(3.69\times10^{-26}\) & \(4.68\times10^{-25}\)\\
\bottomrule
\end{tabular}
\end{table}

\FloatBarrier
\section*{Appendix C---FDTD extraction, floors and external cross-check}
\appendixnumbering{C}

The finite-section readout is computed with a two-dimensional TE finite-difference time-domain scheme following Yee's staggered-grid construction\cite{yee_fdtd_1966}. The dielectric update is written in displacement form,
\begin{equation}
D_y^{\,n+1}=D_y^{\,n}+\Delta t\,(\nabla\times H)^{n+1/2},\qquad
E_y^{\,n+1}=\frac{D_y^{\,n+1}}{\epsilon_0\eps_r(x,z,t_{n+1})}.
\label{eq:app_fdtd_update}
\end{equation}
This is the discrete counterpart of the temporal-interface condition used in the bulk solver: \(D_y\) is the continuous field through the time-varying dielectric. The longitudinal boundaries are terminated by third-order polynomial absorbing layers of thickness \(1.2h\) and maximum damping coefficient 3.5.

The output observable is obtained from the field recorded in the static guide after the section. We project that field onto the static PEC modes and demodulate it with a Hann-windowed Fourier component at \(\omega_m=\omega_0+m\Omega_{\rm mod}\). The steady-state record spans 40 modulation periods after a 32-period transient; this record is the spectral window used for all sideband amplitudes. The spectrally isolated run analyses \(m=0,+1\) for glide-rule floors, while the open-channel run analyses \(m=0,\ldots,+3\). For each propagating sideband,
\begin{equation}
P_{n,m}\propto\frac{\beta_{n,m}}{|\omega_m|}|a_n(\omega_m)|^2,\qquad
\beta_{n,m}=\sqrt{\bar\eps_r(\omega_m/c)^2-(n\pi/h)^2}.
\label{eq:app_fdtd_power}
\end{equation}
Evanescent channels are excluded from the propagating-power sums. The wrong-parity fraction is evaluated sideband by sideband as \(W_m=P_m^{\rm wrong}/(P_m^{\rm wrong}+P_m^{\rm right})\).

\begin{table}[tb]
\centering
\caption{Numerical floors and rule-resolved sideband checks. The synchronous rows measure the forbidden even-sector content of the carrier, while the refined glide rows check that the glide endpoint remains at the same floor. The refined grid uses \((\Delta x,\Delta z)=(0.02h,0.0125h)\).}
\label{tab:app_fdtd_floors}
\small
\begin{tabular}{llcc}
\toprule
configuration & calibration run & analysed sidebands & floor estimate\\
\midrule
spectrally isolated & sync forbidden sector & \(m=0\) & \(3.2\times10^{-13}\)\\
spectrally isolated & refined glide & \(m=0,+1\) & \(1.05\times10^{-14}\)\\
open channel & sync forbidden sector & \(m=0\) & \(1.4\times10^{-11}\)\\
open channel & refined glide & \(m=0,\ldots,+3\) & \(1.0\times10^{-11}\)\\
\bottomrule
\end{tabular}
\end{table}

\begin{table}[tb]
\centering
\caption{Representative FDTD controls for the spectrally isolated converter. The first two rows are the Fig.~\ref{fig:fdtd-scattering} runs. The remaining glide rows are convergence controls with the same target channel; their purpose is to test the parity hierarchy, not to quote an absolute conversion efficiency.}
\label{tab:app_fdtd_controls}
\small
\begin{tabular}{lcc}
\toprule
run & target/generated & \(C_{\rm opp}\)\\
\midrule
Fig.~\ref{fig:fdtd-scattering} glide & 0.901 & 0.930\\
Fig.~\ref{fig:fdtd-scattering} sync & 0 & below floor\\
glide fine grid control & 0.985 & 0.936\\
glide long-window control & 0.981 & 0.968\\
glide shifted-output control & 0.941 & 0.891\\
glide wider-sideband control & 0.977 & 0.967\\
glide longer-ramp control & 0.976 & 0.966\\
\bottomrule
\end{tabular}
\end{table}

\FloatBarrier
The phase sweep of Fig.~\ref{fig:selection-rule} uses the same sideband-resolved quantity, \(W_{+1}\). In the open-channel configuration, \(W_{+1}\) decreases from \(4.81\times10^{-1}\) at \(\phi/\pi=0.5\) to \(6.89\times10^{-4}\) at \(\phi/\pi=0.98\), still many orders above the \(1.38\times10^{-11}\) floor, and then collapses to the floor at \(\phi=\pi\). Fitting the near-glide open-channel points with \(\phi/\pi\ge0.90\) gives
\begin{equation}
W_{+1}\simeq1.7(1-\phi/\pi)^2 .
\label{eq:app_near_glide_fit}
\end{equation}
We use this local fit as an engineering tolerance estimate, not as a model for the full phase sweep.

As an external check, the same scalar TE geometry was reproduced in COMSOL Multiphysics 6.4 using the Electromagnetic Waves, Transient interface. The model used an out-of-plane electric field, PEC plates, the same \(\mathrm{TE}_2\)-like current source and the same sinusoidal modulation phase convention. The fields were exported from \(5\) to \(9.9989\,{\rm ns}\) with a uniform step of approximately \(1.3\,{\rm ps}\), using manual transient time stepping. The synchronous target signal is numerically negligible in the raw out-of-plane field-intensity ratio for the \(\mathrm{TE}_2\rightarrow\mathrm{TE}_3\), \(m=+1\) channel, where the out-of-plane field is \(E_z\) in the COMSOL convention and \(E_y\) in the notation of this paper. This raw COMSOL ratio is used only as an external hierarchy check and is not compared with the calibrated FDTD floors. The glide run places about \(89\%\) of the generated modal power in that target sideband.

\FloatBarrier
\section*{Appendix D---Microwave scales and symmetry-breaking budget}
\appendixnumbering{D}

The finite-section calculation is dimensionless. For a guide height \(h=30\,{\rm mm}\), the Fig.~\ref{fig:fdtd-scattering} parameters correspond to a \(120\,{\rm mm}\) modulated section, \(24\,{\rm mm}\) tapers, \(f_0=8.99\,{\rm GHz}\), \(f_{\rm mod}=3.25\,{\rm GHz}\) and target sideband \(f_{+1}=12.24\,{\rm GHz}\). Other guide heights scale all frequencies as \(1/h\).

\begin{table}[tb]
\centering
\caption{Representative microwave scaling of the finite-section protocol.}
\label{tab:app_microwave_scaling}
\small
\begin{tabular}{lccccc}
\toprule
\(h\) & \(L=4h\) & taper \(0.8h\) & \(f_0\) & \(f_{\rm mod}\) & \(f_{+1}\)\\
\midrule
30 mm & 120 mm & 24 mm & 8.99 GHz & 3.25 GHz & 12.24 GHz\\
50 mm & 200 mm & 40 mm & 5.39 GHz & 1.95 GHz & 7.34 GHz\\
75 mm & 300 mm & 60 mm & 3.60 GHz & 1.30 GHz & 4.90 GHz\\
100 mm & 400 mm & 80 mm & 2.70 GHz & 0.98 GHz & 3.67 GHz\\
\bottomrule
\end{tabular}
\end{table}

\FloatBarrier
The large simulated swing \(\Delta\eps_r=0.65\) makes the converted sideband visible in a short section; the selection rule itself does not require it. In the weak-modulation regime the converted field is linear in modulation depth and the converted power is quadratic. Reducing the effective swing to \(0.10\) or \(0.05\) lowers generated sideband powers by about \(16\) or \(22\) dB relative to the displayed FDTD case, while leaving the parity ratio \(W_m\) fixed by symmetry.

The main experimental tolerance is symmetry breaking. A perturbation that commutes with the glide operation, including glide-symmetric loss or glide-symmetric tapering, changes signal levels but does not populate the forbidden parity sector. A perturbation that breaks the glide relation enters the forbidden-channel amplitude linearly, so \(W_m\) is quadratic in the small error. The phase-delay sweep gives a direct calibration: Eq.~\eqref{eq:app_near_glide_fit} implies that keeping the relative phase within \(\Delta\phi\simeq4^\circ\) of the glide point keeps \(W_{+1}<10^{-3}\). At \(f_{\rm mod}=3.25\,{\rm GHz}\), this phase tolerance corresponds to a timing offset of about \(3.4\,{\rm ps}\) between the two modulation banks.

The corresponding measurement is direct. It can be implemented in a parallel-plate guide with phase-locked tunable inclusions, or in a transmission-line/metasurface-network analogue that discretizes the transverse coordinate. In the network version, \(P_z\) becomes mirror exchange about the centre line and the two outer banks are driven either in phase (synchronous control) or with a \(\pi\) phase delay (time glide). The experimentally robust readout is the complex output profile demodulated at \(f_0+mf_{\rm mod}\), projected onto static guide modes and grouped by static parity.

\end{document}

%% file: figures/fig1_caption.tex
\caption{\textbf{Symmetry structure of synchronous versus time-glide modulation of a trilayer waveguide.} \textbf{a,} Static geometry of the trilayer waveguide. The central slab $|z|<h/6$ has fixed permittivity $\bar{\varepsilon}=1.625$. The two outer slabs, $-h/2<z<-h/6$ and $h/6<z<h/2$, have time-dependent permittivities $\varepsilon_-(t)$ and $\varepsilon_+(t)$ that take values in $\{\varepsilon_1,\varepsilon_2\}=\{1.25,2.0\}$ according to the modulation protocol. The waveguide is bounded by perfect-electric-conductor plates at $z=\pm h/2$. \textbf{b,} In the synchronous protocol both outer layers carry the same permittivity at every instant, so the dielectric profile is reflection symmetric, $\varepsilon(z,t)=\varepsilon(-z,t)$, at all times. The Floquet operator therefore decomposes into static reflection sectors. \textbf{c,} In the time-glide protocol the two outer layers are interchanged after a half period, so the instantaneous profile is not reflection symmetric, but reflection combined with a half-period time translation is a symmetry, $\varepsilon(z,t+T/2)=\varepsilon(-z,t)$. This is the temporal-glide operation. \textbf{d,} Consequence for the harmonic sideband ladder. Synchronous modulation keeps a single static parity across all sidebands. Time-glide modulation does not conserve static parity for the full Floquet mode; instead, it enforces the exact harmonic-wise selection rule $P_z\phi_m=\sigma(-1)^m\phi_m$, where $P_z$ is transverse reflection and $\phi_m$ is the $m$th temporal Floquet harmonic. The allowed parity alternates with sideband index, so transitions between sectors occur only at odd Floquet sidebands and are forbidden at even ones, including the carrier.}